\documentclass[12pt]{iopart}

\usepackage{graphicx}

\begin{document}

\title[The Spin of Supermassive Black Holes]{The Spin of Supermassive Black Holes}

\author{Christopher S. Reynolds}

\address{Dept. of Astronomy and the Joint Space Science Institute, 
University of Maryland, College Park, MD~20901, USA.}
\ead{chris@astro.umd.edu}

\begin{abstract}
Black hole spin is a quantity of great interest to both physicists and astrophysicists.   We review the current status of spin measurements in supermassive black holes (SMBH).   To date, every robust SMBH spin measurement uses X-ray reflection spectroscopy, so we focus almost exclusively on this technique as applied to moderately-luminous active galactic nuclei (AGN).   After describing the foundations and uncertainties of the method, we summarize the current status of the field.   At the time of writing, observations by {\it XMM-Newton}, {\it Suzaku} and {\it NuSTAR}  have given robust spin constraints on 22 SMBHs.  We find a significant number of rapidly-rotating SMBHs (with dimensionless spin parameters $a>0.9$) although, especially at the higher masses ($M>4\times 10^7M_\odot$), there are also some SMBHs with intermediate spin parameters.  This may be giving us our first hint at a mass-dependent spin distribution which would, in turn, provide interesting constraints on models for SMBH growth.   We also discuss the recent discovery of relativistic X-ray reverberation which we can use to ``echo map" the innermost regions of the accretion disk.  The ultimate development of these reverberation techniques, when applied to data from future high-throughput X-ray observatories such as {\it LOFT}, {\it ATHENA+}, and {\it AXSIO}, will permit the measurement of black hole spin by a characterization of strong-field Shapiro delays.    We conclude with a brief discussion of other electromagnetic methods that have been attempted or are being developed to constrain SMBH spin.  
\end{abstract}

\maketitle

\section{Introduction}\label{sec:intro}

Supermassive black holes (SMBHs) are commonplace in today's Universe, being found at the center of essentially every galaxy \cite{kormendy:95a}.  As such, they have attracted the attention of astrophysicists and physicists alike.    Astrophysicists have long been fascinated by the extremely energetic phenomena powered by the accretion of matter onto black holes \cite{rees:84a}.  Beyond the natural draw of studying ``cosmic fireworks", it is now believed that the energy released by growing SMBHs is important in shaping the properties of today's galaxies and galaxy clusters \cite{benson:10a,fabian:12b}.  Thus, SMBHs are important players in the larger story of galactic-scale structure formation.  To the gravitational physicist, SMBHs provide the ultimate laboratory in which to test the prediction of General Relativity (GR) and it's extensions \cite{berti:13a}.

This article discusses recent progress in observational studies of strong gravitational physics close to SMBHs and, in particular, the characterization of black hole (BH) spin.  In order to bring focus to this discussion, we shall address exclusively studies of SMBHs.   A large body of parallel work exists directed towards stellar-mass BHs; we direct the interested reader to some excellent reviews \cite{miller:07a,mcclintock:13a}.     Before beginning our discussion, it is worth reiterating why SMBH spin is an interesting quantity to pursue.  Assuming that standard GR describes the macroscopic physics, spin provides a powerful way to probe the growth history of SMBHs and, ultimately, their formation pathways (an issue that remains mysterious).   In essence, scenarios in which SMBH growth is dominated by BH-BH mergers predict a population of modestly spinning SMBHs, whereas growth via gas accretion can lead to a rapidly-spinning or a very slowly-spinning population depending upon whether the accreting matter maintains a coherent angular momentum vector over the time it takes to double the BH mass \cite{moderski:96a,volonteri:05a}.  SMBH spin can also be a potent energy source, and may well drive the powerful relativistic jets that are seen from many BH systems \cite{blandford:77a,mckinney:12a}.  On more fundamental issues, generic deviations from or extensions to GR tend to show up at the same order as spin effects (e.g. \cite{johannsen:10a}) --- thus observational probes capable of diagnosing spin are also sensitive to physics beyond GR.   In addition, spinning BHs have recently been shown to be unstable in extensions of GR with massive gravitons; for sufficiently massive gravitons, SMBHs should spin down on astrophysically-relevant timescales \cite{berti:13a,brito:13a} (although see \cite{brito:13b}).  Thus, observations of spinning SMBHs can set upper limits on the graviton mass.

We structure this paper as follows.  In Section~\ref{sec:physics}, we briefly review the basic physics relevant to the discussion of accreting SMBHs.  In Section~\ref{sec:x-ray}, we describe the basic anatomy of an accreting SMBH (Section~\ref{sec:basic_structure}), and then present the main tool employed to measure BH spin, namely X-ray reflection spectroscopy (Sections~\ref{sec:reflection} and \ref{sec:relativistic_effects}).  We illustrate the method with a detailed discussion of the SMBH in the galaxy NGC~3783 (Section~\ref{sec:worked_example}), discuss the uncertainties and caveats that should be attached to these spin measurements (Section~\ref{sec:health_warnings}), and then summarize recent attempts to map out the distribution of BH spins (Section~\ref{sec:survey}).  While the long-term future of precision SMBH spin measurements will likely be dominated by space-based gravitational wave astronomy \cite{2013arXiv1305.5720C}(GWA), Section~\ref{sec:conclusions} concludes with a brief discussion of other electromagnetic probes of spin that are complementary to X-ray spectroscopy and can be developed in the pre-GWA era.  

In what follows, the mass of the SMBH shall be denoted $M$.  Following the usual convection, we shall mostly use units in which $G=c=1$; thus, length and time are measured in units of $M$ with $1M_\odot\equiv 1.48\,{\rm km}\equiv 4.93\mu {\rm s}$

\section{Review of the basic physics}\label{sec:physics}

We begin by stating two basic but important facts about a SMBH at the center of a galaxy.  Firstly, the BH is believed to dominate the gravitational potential out to at least $10^5M$.  Secondly, the ubiquitous presence of plasma in the galactic environment precludes any significant charge build-up on the BH.  Assuming standard GR (as we shall do henceforth), these two conditions imply that the spacetime is characterized by the Kerr metric \cite{kerr:63a}, parameterized solely by the mass $M$ and dimensionless spin parameter $a\equiv Jc/GM^2$, where $J$ is the angular momentum of the BH.  Following the usual astrophysical convention, we shall work in Boyer-Lindquist coordinates \cite{boyer:67a} in which the Kerr line element is given by
\begin{eqnarray}
  ds^2&=&-\left(1-\frac{2Mr}{\Sigma}\right)dt^2 -
  \frac{4aM^2r\sin^2\theta}{\Sigma}dt\,d\phi +
  \frac{\Sigma}{\Delta}dr^2\\\nonumber
&+&\Sigma\,d\theta^2+\left(r^2+a^2M^2+
\frac{2a^2M^3r\sin^2\theta}{\Sigma}\right)\sin^2\theta\,d\phi^2,
\end{eqnarray}
where $\Delta=r^2-2Mr+a^2M^2$, $\Sigma = r^2+a^2M^2\cos^2\theta$.  The {\it event horizon} is given by the outer route of $\Delta=0$, i.e., $r_{\rm evt}=M(1+\sqrt{1-a^2})$.

There are two other locations of special astrophysical importance within the Kerr metric.  The surface defined by $\Sigma=0$ (which completely encompasses the event horizon) is the {\it static limit}; within the static limit, any (time-like or light-like) particle trajectory is forced to rotate in the same sense as the BH as seen by a distant observer (i.e. $d\phi/dt>0$).  The stationary nature of the spacetime (i.e. the existence of a time-like Killing vector, $\partial_t$) allows us to define the total conserved energy of any test-particle orbit around the BH.  A curious property of the Kerr metric is the existence of orbits within the static limit (but exterior to the event horizon) for which this conserved energy is negative.  Penrose \cite{penrose:69a,mtw:73} showed that the existence of these orbits permits, in principle, the complete extraction of the rotational energy of a spinning BH, resulting in this region being called the {\it ergosphere}.  Blandford \& Znajek \cite{blandford:77a} demonstrated that magnetic fields  supported by currents in the surrounding plasma can interact with a spinning BH to realize a Penrose process; this is the basis of spin-powered models for BH jets.

A second astrophysically important location is the {\it innermost stable circular orbit} (ISCO), i.e. the radius inside of which circular test-mass (time-like) orbits in the $\theta=\pi/2$ plane become unstable to small perturbations.  This is given by the expression,
\begin{eqnarray}
r_{\rm isco}&=&M\left(3+Z_2\mp \left[(3-Z_1)(3+Z_1+2Z_2) \right]^{1/2} \right),
\end{eqnarray}
where the $\mp$ sign is for test particles in prograde and retrograde
orbits, respectively, relative to the spin of the BH and we have defined,
\begin{eqnarray}
Z_1&=&1+\left(1-a^2\right)^{1/3}\left[\left(1+a\right)^{1/3}+\left(1-a\right)^{1/3}\right],\\
Z_2&=&\left(3a^2+Z_1^2\right)^{1/2}.
\end{eqnarray}
For $a=1$ (maximal spin in the prograde sense relative to the orbiting particle), we have $r_{\rm isco}=M$.   This is the same coordinate value as possessed by the event horizon but, in fact, the coordinate system is singular at this location and there exists finite proper distance between the two locations.  As $a$ decreases, $r_{\rm isco}$ monotonically increases through $r_{\rm isco}=6M$ when $a=0$ to reach a maximum of $r=9M$ when $a=-1$ (maximal spin retrograde to the orbiting particle).  As we discuss below, the ISCO sets an effective inner edge to the accretion disk (at least for the disk configurations that we shall be considering here).   Thus, the spin dependence of the ISCO directly translates into spin-dependent observables; as spin increases and the radius of the ISCO decreases, the disk becomes more efficient at extracting/radiating the gravitational binding energy of the accreting matter, the disk becomes hotter, temporal frequencies associated with the inner disk are increased, and the gravitational redshifts of the disk emission are increased.

\section{X-ray spectroscopic measurements of SMBH spin}\label{sec:x-ray}

We now turn to a discussion of the main observational technique that has been used to date to obtain robust measurements of SMBH spin --- X-ray reflection spectroscopy.   


\subsection{Basic structure of a SMBH accretion disk}\label{sec:basic_structure}

\begin{figure}[t]
\centerline{
\includegraphics[width=0.8\textwidth]{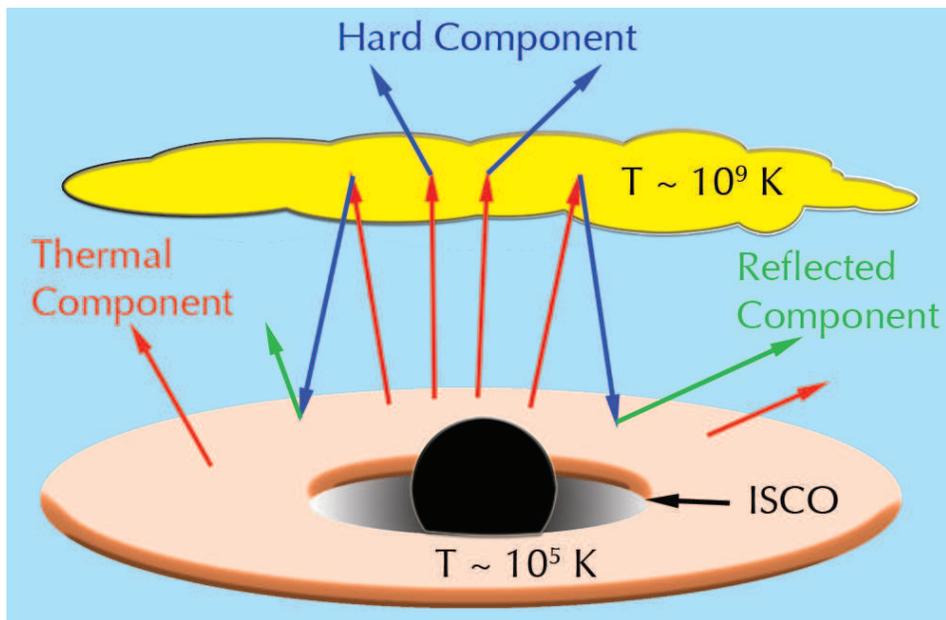}
}
\caption{Schematic sketch of the accretion disk geometry discussed in Section~\ref{sec:basic_structure}.  The accretion disk is geometrically-thin and flat (in the symmetry plane of the Kerr BH).  Within the ISCO, the flow plunges into the BH, effectively truncating the disk signatures at the ISCO.  In SMBH systems, the disk has a temperature of $\sim 10^5$\,K and the X-rays arise from a geometrically extended structure (the ``corona") elevated off the disk plane.  The X-ray source irradiates the accretion disk's surface, producing observable reflection signatures in the X-ray spectrum.  Figure courtesy of Lijun Gou, closely following similar figure in \cite{gou:11a}.}
\label{fig:diskcartoon}
\end{figure}

In order to characterize the rather subtle effects of BH spin, we must select astrophysical systems that we believe possess particularly simple and well-defined accretion geometries.  In our discussion, we only consider sources that have ``moderately high" accretion rates, i.e., those with luminosities in the range $\sim 10^{-2}L_{\rm Edd}$ to $\sim 0.3L_{\rm Edd}$ where $L_{\rm Edd}\approx 1.3\times 10^{31}(M/M_\odot)\,{\rm W}$ is the usual Eddington limit (the luminosity above which fully ionized hydrogen is blown away from the BH by radiation pressure).  This means that we are studying bonafide active galactic nuclei (AGN).    For such systems, the accretion geometry is described by the cartoon shown in Fig.~\ref{fig:diskcartoon}.  Outside of the ISCO, the accretion disk is geometrically-thin and consists of gas following almost test-particle circular orbits  (hereafter, we shall use the term ``keplerian motion" to mean test-particle circular orbits).  The gradual re-distribution of angular momentum by magnetohydrodynamic (MHD) turbulence within the disk \cite{shakura:73a,balbus:91a} leads to a gradual inwards radial drift of matter superposed on this background of circular motion.  Inside of the ISCO, material spirals rapidly into the BH approximately conserving energy and angular momentum.  The radial velocities become relativistic (becoming $c$ at the horizon itself) and so, by conservation of baryon number, the density of the flow plummets within this region, and the matter becomes fully ionized.  Indeed, it may even become optically-thin as indicated in the cartoon.  The transition of the accretion flow from approximately circular motion to inward plunging at the ISCO is particularly important for our discussion --- it sets an  inner radius to the region of the disk that produced any observable effects..  This transition is seen in MHD simulations of accretion disks \cite{reynolds:08a,penna:10a}, although a full characterization of the inner disk structure must await the development of global, radiation-dominated GR-MHD simulations.  Noting that accretion disk theory is still an active and on-going field of research, we shall proceed under the assumption that this transition occurs at the ISCO (although, as we shall point out when appropriate, small shifts in the transition away from the ISCO due to finite disk thickness effects may well be an important source of systematic error for BH spin).  

In general, of course, the angular momentum vector of the incoming material about the BH (and hence the orientation of the accretion disk at large radius) may be misaligned with the spin angular momentum of the BH.  In what is now the standard picture, Bareen \& Peterson \cite{bardeen:75a} argue that  the effects of differential Lens-Thirring precession slave the inner accretion disk to the BH symmetric plane ($\theta=\pi/2$), producing a disk that gradually twists from alignment with the mass reservoir at the outside, to alignment with the black hole in its inner region.  The Bardeen-Petterson argument suggests that the inner $r\sim 100M$ of the accretion disk will be essentially flat in the $\theta=\pi/2$ plane.  However, the details of this process, including the value of the characteristic warp radius, depends upon the transport of the out-of-plane angular momentum in an MHD turbulent disk; only now has it become tractable to examine this problem with MHD simulations and we eagerly await results.  For the rest of the discussion, the effects of disk warping shall be neglected.

The temperature of the inner accretion disk around a SMBH for these accretion rates is generally in the range $10^5-10^6\,K$; the resulting thermal (quasi-blackbody) emission produces a strong bump peaking in the UV/EUV.   Indeed, this UV bump can often dominate the observed energy output of an AGN.  However, essentially every AGN also shows significant emission in the X-ray band.  This has the form of a power-law component with flux-density $F(\nu)\propto \nu^{-\alpha}$, $\alpha\approx 1$, and can carry up to 10--20\% of the total power radiated by the AGN. This ``hard component" emission must arise from much hotter, optically-thin material associated with either a magnetically heated ``corona" above the disk or dissipation in the base regions of a jet (the yellow cloud in Fig.~\ref{fig:diskcartoon}).  On the basis of detailed spectral studies, the most likely physical process producing the X-ray emission is the inverse Compton scattering of optical/UV photons from the accretion disk by $\sim 10^9\,{\rm K}$ electrons in the corona.  

\subsection{X-ray ``reflection" from the SMBH accretion disk}\label{sec:reflection}

This configuration of a flat, thin, cold, keplerian disk below a strong continuum X-ray source proves to be fortuitous.   The observed X-ray emission is (usually) dominated by radiation that comes straight from the corona to us.  However, the X-rays from the corona also irradiate the accretion disk.   With  energy-dependent ratios, the incident photons are either photoelectrically-absorbed by ions in the surface layers of the disk or Compton scattered back out of the disk.  The de-excitation of the resulting excited ions produces a rich X-ray spectrum of radiative-recombination features and fluorescent emission lines.  Collectively, the Compton scattered continuum and these emission features make up (what is poorly termed) the {\it X-ray reflection spectrum} \cite{lightman:88a,george:91a,ross:05a,garcia:13a}.

\begin{figure}[t]
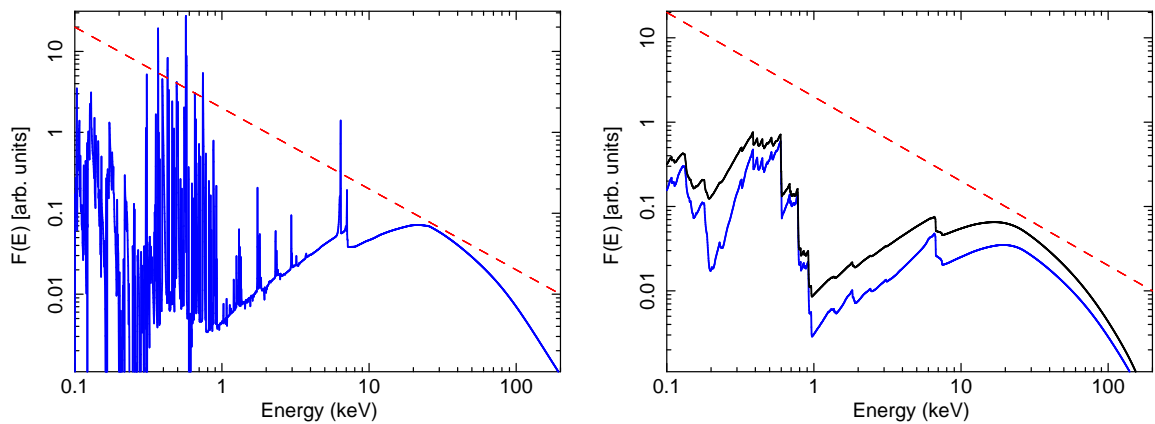

\centerline{
\includegraphics[width=0.35\textwidth,angle=-90]{f2a.ps}
\includegraphics[width=0.35\textwidth,angle=-90]{f2b.ps}
}
\caption{{\it Left panel : }Incident power-law continuum (red dashed line) and rest-frame reflection spectrum produced by the accretion disk (solid blue line) computed using the {\tt xillver} model of \cite{garcia:13a} assuming an ionization parameter of $\xi=5\,{\rm erg}\,{\rm cm}\,{\rm s}^{-1}$ and cosmic iron abundance.  {\it Right panel : }Effects of relativistic Doppler and gravitational redshift effects on this reflection spectrum assuming a viewing inclination of $i=30^\circ$, irradiation index of $q=3$, and a rapidly spinning BH ($a=0.99$; black line) or non-spinning ($a=0$; blue line) BH.}
\label{fig:reflection}
\end{figure}

A representative calculation of the rest-frame X-ray reflection (using the {\tt xillver} code of \cite{garcia:13a}) is shown in Fig.~\ref{fig:reflection} (left).   Particularly important is the iron-K$\alpha$ line at 6.4\,keV \cite{fabian:89a}.  This line is strong due to the high astrophysical abundance and high fluorescence yield of iron.  It is also relatively isolated in the spectrum, making it easy to identify and characterize even if it is strongly broadened and distorted (as we will discuss in Section~\ref{sec:relativistic_effects}).  However, there are other characteristic features of the reflection spectrum.  At low X-ray energies, the spectrum is dominated by a dense forrest of emission lines from low atomic number elements (with N, O, Ne, Mg, Si, and S being particularly important).  At high X-ray energies, the absorption cross-sections become small and Compton-scattering dominates, producing a broad hump in the spectrum that peaks at 20--30\,keV.  Of course, the precise form of the spectrum depends upon several things, most notable (i) the ionization state of the disk surface (characterized by the ionization parameter $\xi=4\pi F_{\rm ion}/n$ where $F_{\rm ion}$ is the ionizing flux impinging on the disk surface and $n$ is the electron number density at the X-ray photosphere), (ii) the elemental composition of the disk matter (with the iron fraction being of special importance), and (iii) the shape of the irradiating continuum.

\subsection{Relativistic effects and spin-effects in the X-ray spectrum}\label{sec:relativistic_effects}

The geometry outlined above is a relativists dream --- the accretion disk is a thin structure following almost test-particle orbits close to the BH and emitting well-defined, sharp spectral features.  The observed spectrum will thus be strongly modified by relativistic Doppler shifts and gravitational redshift (see Fig.~\ref{fig:reflection} [right]).   By identifying these spectral features and modeling the broadening effects we can determine the properties of the accretion disk itself (e.g., the ionization state of the surface layers and the elemental abundances of the disk) as well as the inclination of the disk and the location of the ISCO (hence the BH spin).

Operationally, the analysis of real data is facilitated by making the approximation that the rest-frame disk reflection spectrum is invariant across the region of interest.  We then model the observer-frame spectrum by convolving the reflection spectrum with a kernel describing the GR photon transfer effects (Doppler and gravitational shifts).   This kernel is constructed by ray-tracing from the the disk surface ($\theta=\pi/2$ plane) to the observer, taking full advantage of the conserved quantities in the Kerr metric (see \cite{dauser:10a} for a review of this calculation).   The kernel depends upon the viewing inclination of the accretion disk and the BH spin.     It also depends upon the radial distribution of the irradiating X-ray flux (and hence the weight that should be given to each radius of the disk in the blurring kernel); this is an unknown function and something we would like to determine from the data.  Thus, these kernels are calculated and tabulated as a function of emission radius, and the full kernel is constructed (for a given viewing inclination and spin) by integrating over radii weighted by the irradiation profile \cite{cunningham:75a,fabian:89a,laor:91a,brenneman:06a}.  We commonly parameterize the irradiation profile with a power-law form, $F_{\rm ion}\propto r^{-q}$, where $q$ is known as the {\it irradiation index}.  High-quality datasets sometimes require a broken power-law form in order to fully capture the shape of the broadened reflection spectrum in which case we have two irradiation indices $(q_1,q_2)$ and a break radius \cite{fabian:02a,brenneman:11a, fabian:12c}.  A non-parametric treatment of the irradiation profile \cite{wilkins:11a} confirms the appropriateness of a broken power-law form.

\subsection{A worked example : the spin of the SMBH in NGC3783}\label{sec:worked_example}

In principle, the procedure to determine SMBH spin is now straightforward.   We take a high signal-to-noise X-ray spectrum of an AGN with a high throughput observatory such as {\it XMM-Newton}, {\it Suzaku} or {\it NuSTAR}.  We then compare model spectra (such as illustrated in Fig.~\ref{fig:reflection} [right]) with the observed spectrum and, using standard techniques (e.g. direct $\chi^2$-minimization or Monte Carlo Markov Chain [MCMC] methods), determine the best-fitting  values and error ranges on all model parameters, including spin.

\begin{figure}[t]
\centerline{
\includegraphics[width=0.55\textwidth,angle=-90]{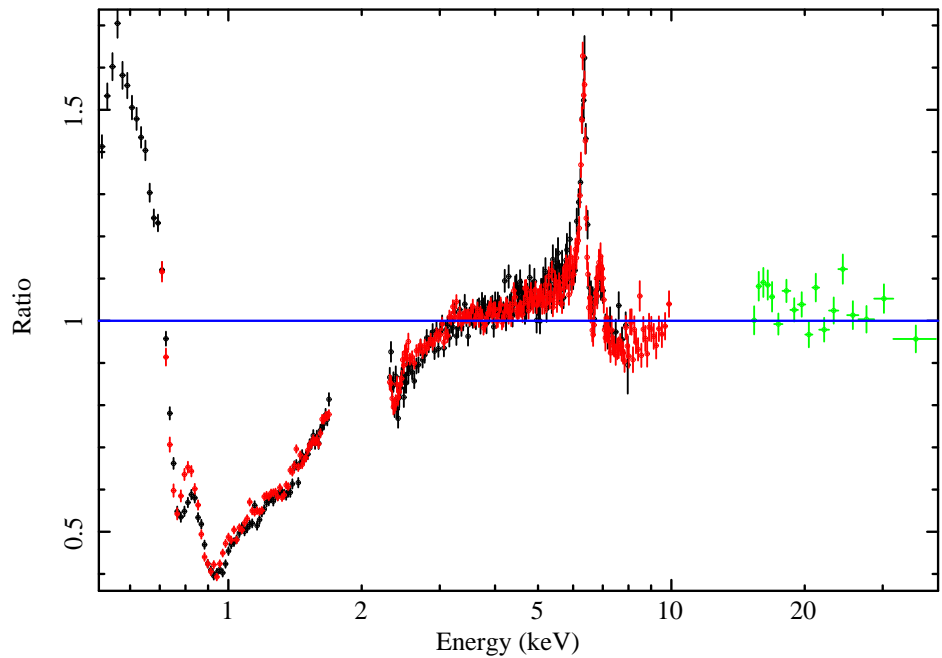}
}
\caption{Th 2009 {\it Suzaku} spectrum of NGC3783 ratioed against a power-law continuum.  Figure from \cite{reis:12a}.}
\label{fig:ngc3783_ratio}
\end{figure}

However, AGN possess complex structures in addition to the accretion disk, including winds of photoionized plasma that are blown off the disk as well as large-scale cold structures that may act as the reservoir for the accretion disk.  These structures imprint their own patterns of absorption, emission and reflection on the X-ray spectrum and these must be included in our spectral model.   We illustrate this with a discussion of the AGN in the spiral galaxy NGC3783, one of the targets of the {\it Suzaku} Key Project on BH spin.  Figure~\ref{fig:ngc3783_ratio} shows the ratio of the {\it Suzaku} data to a reference power-law model.  The soft X-ray spectrum is dominated by an absorption feature from a highly-subrelativistic wind of photoionized plasma flowing away from the BH (the so-called {\it warm absorber}, \cite{halpern:84a,reynolds:97a,netzer:03a,mckernan:07a}).  This absorption can be accurately modeled by spectral synthesis codes such as {\tt xstar} \cite{kallman:01a}.  Shifting attention to the iron-K band around 6\,keV, we see in Fig.~\ref{fig:ngc3783_ratio} a strong narrow iron line due to the fluorescence of cold gas that is distant from the SMBH.   This can be accurately described by a low-ionization X-ray reflection spectrum.

\begin{figure}[t]
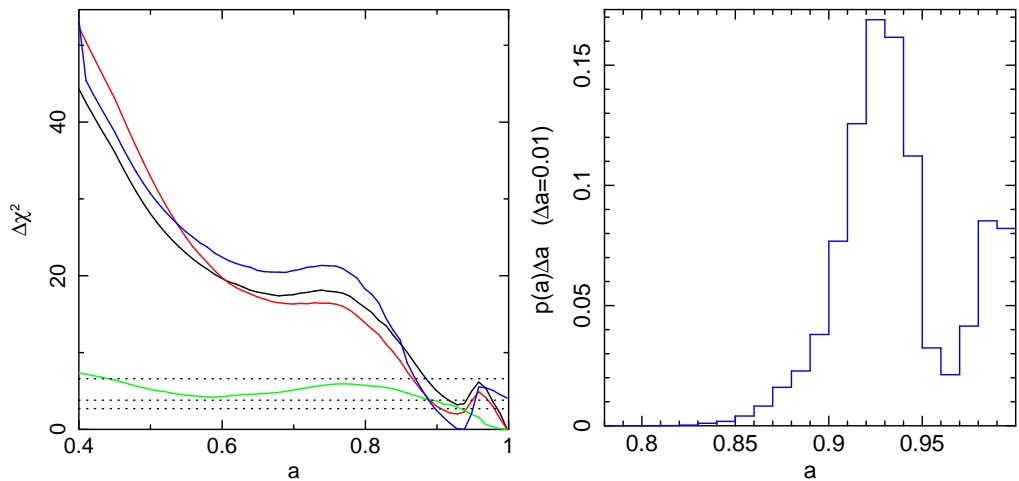

\centerline{
\includegraphics[width=0.4\textwidth,angle=-90]{f4a.ps}
\includegraphics[width=0.4\textwidth,angle=-90]{f4b.ps}
}
\caption{Spin constraints on the SMBH in NGC3783 from the 2009 {\it Suzaku} observation.   {\it Left panel : } $\Delta\chi^2$ as a function of the spin parameter $a$.  Different lines show the effects of different data analysis assumptions; a fiducial analysis (black), an analysis in which the warm absorber parameters are frozen at their best values (red), an analysis in which the cross normalization of the two {\it Suzaku} instruments (the XIS and HXD/PIN) are allowed to float (blue), and an analysis that ignores all data below 3\,keV (green).   Figure from \cite{brenneman:11a}.    {\it Right panel : }Probability distribution for spin parameter $a$ from a Monte Carlo Markov Chain (MCMC) analysis using the fiducial spectral model.  Figure from \cite{reynolds:12a}.}
\label{fig:ngc3783_spinconstraints}
\end{figure}

\begin{figure}[t]
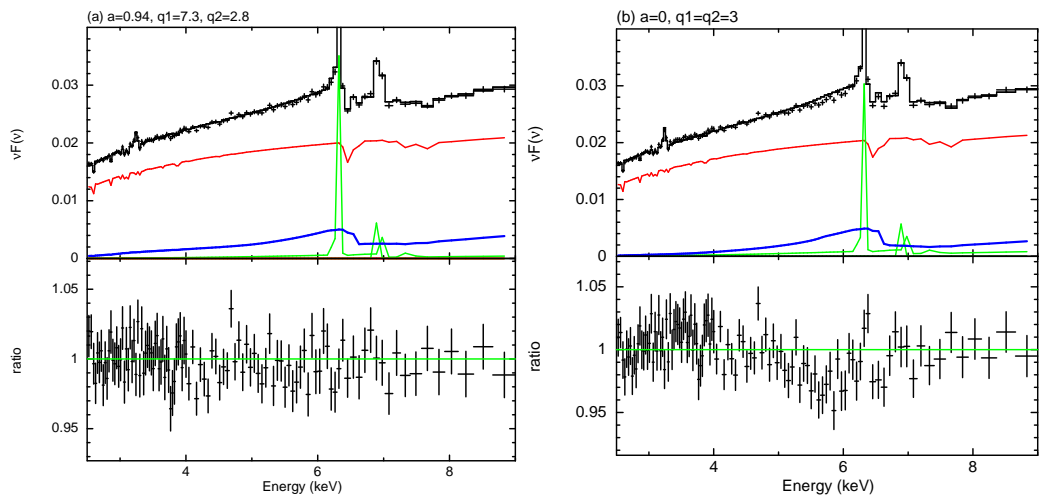

\centerline{
\includegraphics[width=0.42\textwidth,angle=-90]{f5a.ps}
\includegraphics[width=0.42\textwidth,angle=-90]{f5b.ps}
}
\caption{{\it Left panel : }{\it Suzaku}/XIS spectra overlaid with the best fitting model (top) and the corresponding data/model ratio (bottom).  {\it Right panel : }Same, except that the spin parameter has been frozen at $a=0$ and (for physical consistency) the irradiation indices have been frozen at $q_1=q_2=3$.  All other parameters have been allowed to fit freely.   For both panels, the model components are colored as follows: absorbed power-law continuum (red), distant reflection (green), and relativistically smeared disk reflection (blue).      Figure from \cite{reynolds:12a}.}
\label{fig:ngc3783_residuals}
\end{figure}

Our final spectral model must include all of these additional components (with their associated parameters), and the spin constraints must be derived marginalizing over all other parameters.   Figure~\ref{fig:ngc3783_spinconstraints} shows the constraints for NGC3783 derived by both a simple $\chi^2$ minimization procedure and an MCMC approach.  Both approaches constrain the spin to be $a>0.89$ at the 90\% confidence level \cite{brenneman:11a,reynolds:12a}.  Figure~\ref{fig:ngc3783_residuals} illustrates precisely how the models are constraining the spin by showing the residuals between the model prediction and the data when the ``wrong" spin is forced into the fit.  We see that, when a non-spinning BH is imposed in the fit to the NGC3783 data, the broad iron line is too narrow and over-predicts the flux in the 5--6\,keV band.   For further detailed discussion of the statistical robustness of these fits, including a treatment of the subtle statistical correlation between BH spin and iron abundance, see \cite{reynolds:12a}.

\subsection{Health warnings}\label{sec:health_warnings}

As in any field of experimental physics, we must be aware of the systematic errors in our measurements or, worse, the possible ways that our methodology may completely fail to capture the reality of a given astrophysical source.  Here, we walk through some of the main concerns relevant for SMBH spin measurements.  

The single biggest concern is that, in some sources, we may be fundamentally mis-characterizing the spectrum.  In \cite{miller:08a}, it is shown that multiple absorbers that {\it partially} cover the X-ray source can mimic the relativistically-blurred reflection spectrum, at least in the 0.5--10\,keV band.  A significant issue faced by these models is their geometric plausibility --- the absorbers are postulated to be some significant distance from the SMBH ($r\gg100M$) whereas the X-ray source is very compact, lying within $10M$ of the SMBH (the compactness of the X-ray source has recently been confirmed by gravitational microlensing studies of multiply-lensed quasars, e.g. \cite{morgan:12a}).  Thus, partial covering of the X-ray source requires either a very special viewing angle (so that our line of sight just clips the edge of the absorber) or a finely clumped wind with tuned clumping factors (such that, at any given time, $\sim 50$\% of our sight-lines to the compact source are covered by clumps).  Still, it is important to address this model from a hard-headed data point of view.  This was recently achieved using a joint {\it XMM-Newton/NuSTAR} observation of the AGN in the spiral galaxy NGC~1365 that, for the first time, allowed a high-quality X-ray spectrum of an AGN to be measured in the 0.5--60\,keV band.  As reported in \cite{risaliti:13a}, the canonical partial-covering absorption model can fit the 0.5--10\,keV spectrum but fails when extrapolated up to the higher-energy portion of the spectrum.  Adding in an extra component partial covering component that is marginally optically-thick to Compton scattering can produce agreement with the observed spectrum, but then the inferred absorption-corrected luminosity of the AGN would exceed the Eddington limit and hence this solution would be unphysical.  It was concluded by \cite{risaliti:13a} that the data strongly favor the relativistic disk reflection model over partial-covering absorption.  

Within the disk reflection paradigm, there are some important issues to note.  Firstly, our method assumes that the accretion disk observables truncate at the ISCO; the spin constraint is almost entirely driven by this fact.  While there is support for this assumption from MHD simulations of accretion disks \cite{reynolds:08a,penna:10a}, our knowledge of disk physics is incomplete.  In particular, the inner accretion disk may be subject to visco-thermal instabilities that could, sporadically, truncate disk outside of the ISCO.    Although there is no evidence for this transitory disk truncation in most AGN, there is an important class of AGN that do appear to show this phenomenon -- the broad-line radio galaxies (BLRGs).  BLRGs are in our target range of accretion rates, display the strong optical/UV/X-ray emission characteristic of a geometrically-thin accretion disks and corona, but also possess powerful relativistic jets.  The jet experiences abrupt and sporadic injections of energy (i.e., geyser-like eruptions) and, during one of these injection-events, the inner regions of the thin-disk appear to be destroyed before refilling back down to the ISCO\cite{marscher:02a,chatterjee:09a,chatterjee:11a,lohfink:13a}.  Thus, some monitoring of the object is required to know whether the spin measurement can be considered robust or not \cite{lohfink:13a}.

Secondly, there is the question of which aspect of the data is driving the fit.  In the most robust cases, such as NGC3783 discussed in Section~\ref{sec:worked_example}, the constraint is clearly driven by the profile of the broad iron line.  However, in addition to the broad iron line, the ionized reflection spectrum also contributes to the soft part of the spectrum ($<1$\,keV); the forrest of emission features in the soft reflection spectrum are broadened into a pseudo-continuum that produces a rather smooth ``soft excess" in the spectrum.   Many sources do indeed display a soft excess and, in some cases, the high signal-to-noise in the soft part of the spectrum can drive the constraints on the spin parameter. If the observed soft excess is truly due to disk reflection, this is a valid and powerful use of the data.  However, there are other possible origins for the soft excess, most notably inverse Compton scattering of UV photons from the cold disk by ``warm" material (possibly a ``chromosphere" lying on the disk surface).  Thus, in some sources, the origin of the soft excess and the value of the spin parameter are coupled questions that are still under examination (see detailed discussion of this issue for the luminous AGN Fairall~9 in \cite{lohfink:12b}).  

\subsection{Survey of current results}\label{sec:survey}

Without further ado, we present the current state of the field of SMBH spin measurements.  Over the past few years there have been an explosion in the number of published SMBH spin measurements, including samples by Walton et al. \cite{walton:13a} and Patrick et al. \cite{patrick:12a}.  Spins have been published by multiple groups and, in the same timeframe, the spectral models and methodologies have been improved.  Thus, with some hindsight, one cannot simply harvest SMBH spin measurements from the literature without exercising caution.  Following \cite{reynolds:13a}, we propose filtering the published values with quality-control criteria; (i) the spin measurement is based on spectral models employing full disk reflection spectra and not just isolated ``pure iron line" models, (ii) the iron abundance is treated as a free parameter (required since iron abundance and spin are correlated variables), (iii) the inclination is treated as a free parameter and can be constrained (this helps protect against fits that are driven purely by the soft excess), (iv) the irradiation index can be constrained and fits to give $q>2$ (required for a well-posed model in which the reflection spectrum is not unphysically dominated by the outer disk radius).   These filter criteria eliminate 13 sources from the Walton \cite{walton:13a} and Patrick \cite{patrick:12a} samples; the excluded sources span the range of AGN type and luminosity.     Taking the compilation described in Table~1 of \cite{reynolds:13a} and updating with the {\it XMM-Newton/NuSTAR} result on NGC1365 \cite{risaliti:13a} and the multi-epoch study of 3C120 \cite{lohfink:13a}, we now have spin constraints on 22 AGN that satisfy the quality control criteria.

\begin{figure}[t]
\centerline{
\includegraphics[width=0.8\textwidth]{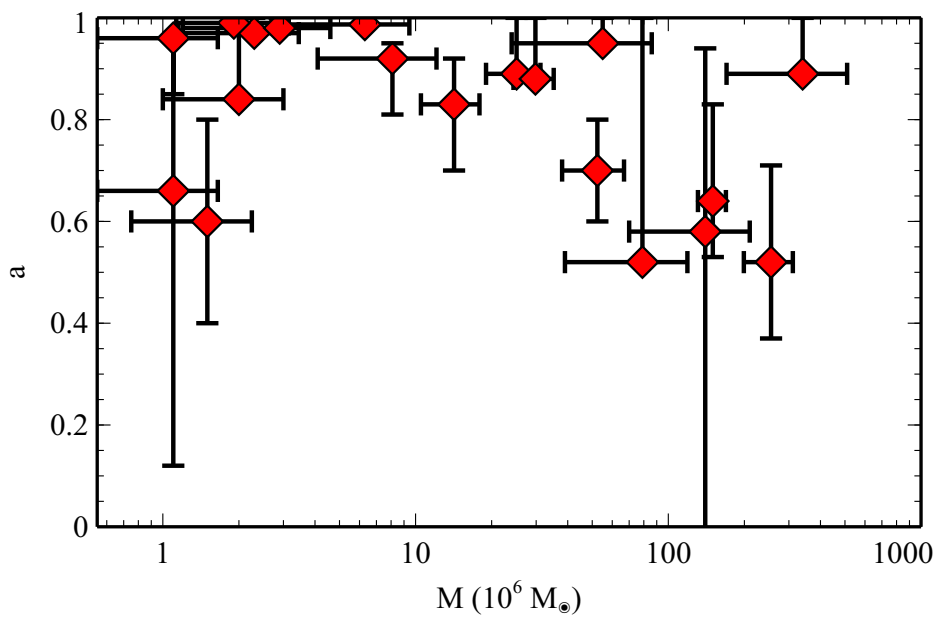}
}
\caption{Masses $M$ and spin parameters $a$ of the 19 SMBHs for which both parameters are constrained.  Following the conventions of the primary literature, the spin measurements are shown with 90\% error ranges, whereas the masses are shown with $1\sigma$.   This is an updated version of a similar figure appearing in \cite{reynolds:13a}.}
\label{fig:smbh_spins}
\end{figure}

In Fig.~\ref{fig:smbh_spins} we take the 19 of these objects that also have mass estimates (from various techniques; see \cite{reynolds:13a}) and place them on the $(M,a)$-plane.  There are several interesting points to note about this plot.  Firstly, there is clearly a population of rapidly spinning BHs ($a>0.9$), especially below masses of $4\times 10^7M_\odot$.  This is a strong indication that these SMBHs grew (at least in their final mass doubling) by the accretion of gas with a coherent angular momentum.   Secondly, there are some SMBHs for which intermediate spins ($0.4<a<0.8$) are inferred, and these tend to be the highest mass systems ($M>4\times 10^7M_\odot$).   While the small number statistics and ill-defined selection effects prevent firm conclusions from being drawn, this may be the first hints for a mass-dependence to the SMBH spin distribution, with a more slowly spinning population (corresponding to growth via BH-BH mergers or incoherent accretion \cite{king:04a}) emerging at the highest masses.  Lastly, there are no retrograde spins measured even though our technique is capable of finding them.  A single epoch analysis of the BLRG 3C120 suggested an accretion disk truncated at $r\sim 10M$, possibly indicating a rapid retrograde spin \cite{cowperthwaite:12a}, but a multi-epoch analysis revealed that this was a rapidly-rotating prograde BH with a disk that undergoes transitory truncation related to jet activity \cite{lohfink:13a}.

\subsection{The emerging field of broad iron line reverberation}

\begin{figure}[t]
\centerline{
\includegraphics[width=0.6\textwidth,angle=-90]{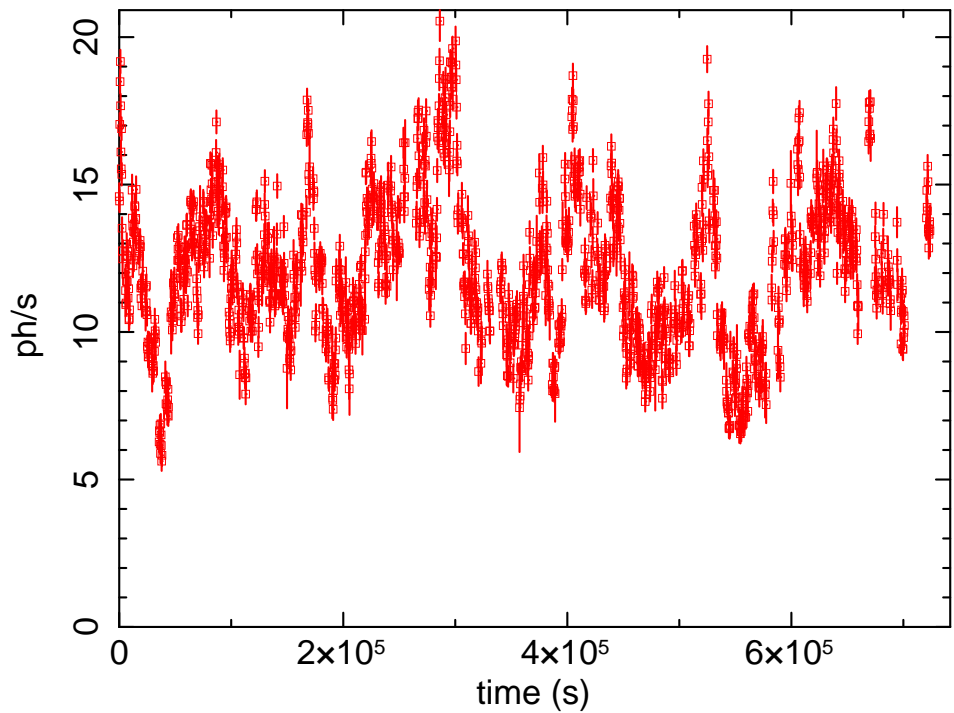}
}
\caption{The 2--10\,keV {\it Rossi X-ray Timing Explorer} light curve of the AGN MCG--6-30-15 from a long observation in August 1997.}
\label{fig:mcg6_rxte}
\end{figure}

An important characteristic of accretion onto BHs that we have not yet addressed is the time-variability.  Fundamentally, the variability is driven by a combination of local instabilities (such as the magnetorotational instability that drives MHD turbulence \cite{balbus:91a}) and more poorly understood global instabilities (that may drive limit-cycle or explosive behavior).  In AGN, this variability is particularly dramatic in the X-ray band since this emission comes from the central regions of accretion disk (where all of the characteristic timescales are shortest) and originates from a corona which may be unstable to, for example, reconnection-driven eruptions.  An example X-ray light curve from the {\it Rossi X-ray Timing Explorer (RXTE)} is shown in Fig.~\ref{fig:mcg6_rxte}.

\begin{figure}[t]
\centerline{
\includegraphics[width=0.9\textwidth]{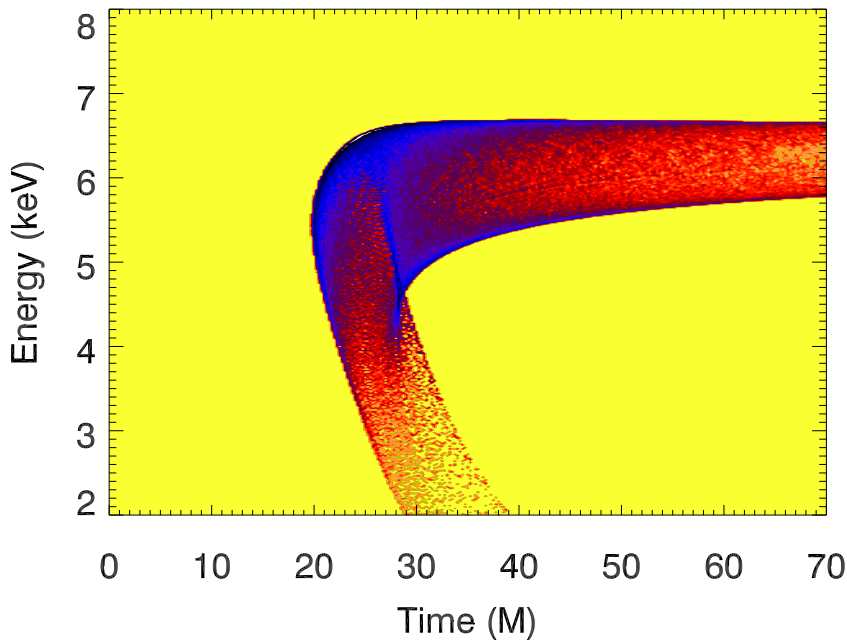}
}
\caption{Relativistic transfer function showing the response of a 6.4\,keV iron line from the disk surface to a $\delta$-function flare in the driving X-ray continuum source.  For the purposes of this illustration, the continuum source is assume to be a point source on the spin axis at $r=10M$, the BH is rapidly spinning $a=0.998$ and the accretion disk has an inclination of $i=30^\circ$. }
\label{fig:reverb_tf}
\end{figure}

Since the X-ray source is elevated off the disk surface, we can use this rapid variability to ``echo map" the reflection from the inner accretion disk.  In essence, by characterizing the time-delays between flares in the direct X-ray continuum and the reflection spectrum, we can map out the geometry of the SMBH/disk/corona system.   This was examined theoretically by the author some years ago \cite{reynolds:99a} who computed transfer functions relating the observed response of the iron-K$\alpha$ emission line from the disk to a $\delta$-function continuum flare.  An example of such a transfer function for a rapidly spinning BH ($a=0.998$) is shown in Fig.~\ref{fig:reverb_tf}.  Beyond the obvious fact that the time-delay gives information on the actual location of the X-ray source (something which is hard to obtain via a pure spectral analysis), it is interesting to note that the transfer function bifurcates into two branches.  The upper branch corresponds to the normal echo sweeping out to larger radii (and irradiating more slowly moving material) giving an ever narrowing line.  The lower branch corresponds to strongly Shapiro-delayed \cite {shapiro:64a} reflection from the very innermost regions of the disk close to the event horizon.  The slope of this branch is a function of the spin parameter $a$, especially when the BH is close to extremal.

While our observatories do not yet have the collecting area to allow us to see this response of the iron line to a single flare, reverberation has recently been discovered via statistical analyzes of long {\it XMM-Newton} datasets of bright AGN.  The analysis is subtle and must account for the fact that (i) there is no single time-delay that characterizes the response of the reflection continuum and, in fact, the measured time delays are both a function of energy and the temporal-frequency being probes; and (ii) the direct continuum source itself has intrinsic energy- and temporal-frequency dependent time lags.  Most analyses use Fourier methods.  For two given (non-overlapping) energy bands, time-sequences of X-ray flux ($a(t)$ and $b(t)$) are extracted and their Fourier transforms ($\tilde{a}(\omega)$ and $\tilde{b}(\omega)$ )are computed.  We then form the cross-spectrum $C(\omega)\equiv \tilde{a}^*\tilde{b}$, which we can write in terms of a frequency-dependent amplitude and phase, $C(\omega)=|C(\omega)|e^{i\phi(\omega)}$.  The frequency-dependent time-delay between the two lightcurves is then given by $\Delta t=\phi(\omega)/\omega$.  

\begin{figure}[t]
\centerline{
\includegraphics[width=0.9\textwidth]{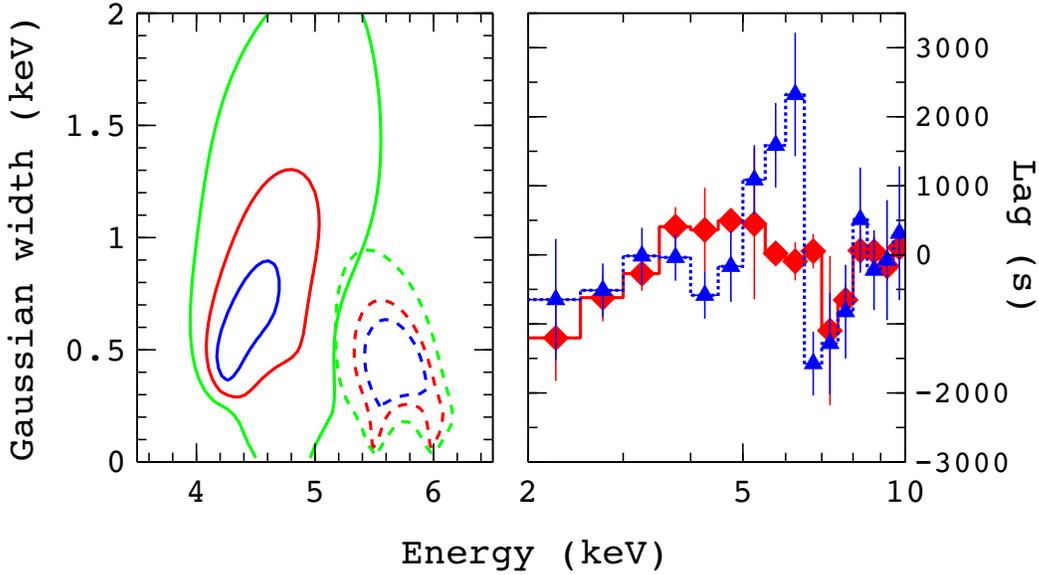}
}
\caption{{\it Right panel : }Time-lags as a function of energy (i.e. the lag-spectrum) for the {\it XMM-Newton} data of the bright AGN NGC4151.  Show here is the lag-spectrum for temporal frequencies $<2\times 10^{-5}$\,Hz (blue triangles) and $(5-50)\times 10^{-5}$\,Hz (red diamonds).  {\it Left panel : }Simple Gaussian fits to these lack spectra, demonstrating that the iron-line feature in the lag spectrum has a centroid energy that decreases and (marginally) a width that increases as one probes high temporal frequencies.  Figure from \cite{zoghbi:12a}.}
\label{fig:ngc4151}
\end{figure}

The first detections of reverberation from the inner accretion disk were made by probing delays between the continuum and the soft excess \cite{fabian:09a,zoghbi:10a,demarco:13a}.  However, recently, there are clear detections of reverberation associated with the broad iron line \cite{zoghbi:12a,zoghbi:13a,kara:13a}.  Figure~\ref{fig:ngc4151} shows the first such detection of broad iron line reverberation in the bright AGN NGC4151.  The magnitude of the lags suggests that the X-ray source is displaced $\sim 4M$ above the center of the disk; this is not necessarily a natural location for strong coronal emission and may suggest that the X-ray source is actually associated with the dissipative base of a jet or some other dissipative structure associated with a BH magnetosphere.  The exact shape of the lag-spectrum depends upon the temporal frequency being probes, with the line appearing broader and more highly redshifted at higher temporal-frequencies.  This is expected since the higher temporal-frequencies are probing smaller radii in the accretion disk, and suggests that we are gaining our first hints at the nature of the transfer function.  

One important point to note (in the spirit of a sanity check) is that the size/location of the X-ray source as inferred from X-ray reverberation is broadly consistent with recent estimates derived from a completely different technique, studies of microlensing in multiply lensed quasars \cite{chartas:09a,dai:10a}. 

\section{Conclusions}\label{sec:conclusions}

In this review, we have summarized the methodology by which we can measure SMBH spins using X-ray reflection spectroscopy and have described the current status of such measurements.  Today, we have spins for approximately two dozen SMBHs; we find a population of rapidly rotating BHs ($a>0.9$) but hints of a more slowly rotating population emerging at higher SMBH masses.  The main limitation to applying this technique more widely are the long integration times needed with current observatories in order to obtain spectra with sufficient signal-to-noise.  The next generation of proposed high throughput X-ray observatories such as {\it LOFT}, {\it ATHENA+} and {AXSIO} will permit us to push these studies to fainter and more distant objects.   This, in turn, will allow well-defined samples to be obtained and permit investigations of the SMBH spin distribution as a function of cosmic time.   These high-throughput observatories will also unlock the full power of reverberation studies, allowing us to detect strong-field Shapiro delays in the reverberation transfer function.

The long term future of SMBH spin measurements will likely be dominated by space-based GWA, once such observatories are deployed.   GWA will permit SMBH spin measurements with a precision and accuracy probably unattainable with X-ray spectroscopy.   However, given the uncertain timeframe for the deployment of space-based GW observatories, it is interesting to pursue electromagnetic probes of SMBH spin as vigorously as possible.   We end this review by mentioning a few other electromagnetic techniques that have been discussed or are actively being developed.

If the accretion disk is efficient at radiating away the gravitational potential energy of the accreting matter, the overall efficiency of the accretion disk (defined as $\eta\equiv L/\dot{M}c^2$, where $L$ is the total radiative luminosity of the system and $\dot{M}\equiv dM/dt$) is a function of spin. With some work, the total $L$ can be estimated straightforwardly.   The mass accretion rate $\dot{M}$ is more challenging to constrain, but it can be estimated through the detailed temperature/shape of the thermal emission from the accretion disk.  Applying this method to a sample of 80 quasars, \cite{davis:11a} found tentative evidence for efficiencies (and hence spin parameters) that increase with increasing SMBH mass.  Of course, this is discrepant with the result presented in Fig.~\ref{fig:smbh_spins} --- future explorations of this discrepancy must properly account for  selection effects as well as the systematic errors in the efficiency-based measurements.

One interesting technique focuses on powerful jetted systems.  If we assume that the jets are powered by the Blandford-Znajek (i.e. magnetic Penrose) process, and if we can make an estimate of the magnetic field in the immediate vicinity of the BH (from accretion theory), then the BH spin parameter can be determined once the jet power is measured.  Practical difficulties include the significant uncertainty in deriving jet powers from observables (which may be strongly influenced by an unknown relativistic beaming factor).  However, it is encouraging that a careful application of this method \cite{daly:11a} produces results that are sensible; given the analytic approximation used to relate spin and jet power, it would be possible to infer $a>1$, but the observed population of AGN seem to obey the constraint $|a|<1$.

Of course, a tremendous amount of observational firepower (including a recent $3$\,Ms campaign by {\it Chandra}) has been directed at the SMBH at the center of our own Galaxy (coincident with the radio source Sgr~A$^*$).  Unfortunately, the X-ray reflection technique cannot be applied here --- the accretion rate is extremely small, and the disk is believed to be in a hot, geometrically-thick (quasi-spherical) state that is optically-thin to X-ray emission and hence does not produce any X-ray reflection features \cite{rees:82a,narayan:95a}.   Instead, calculations of photon production/propagation through GR-MHD simulations of the hot flow have been used to make predictions of the radio spectrum and frequency-dependent polarization as a function of system parameters.  In principle, comparison with the observations can then constrain the spin.  However, due to the inherent uncertainties in the calculation of this complex radio continuum, the constraints to date have not yet proved conclusive, with some groups obtaining high inferred spin \cite{moscibrodzka:09a,dexter:11a} and others obtaining rather lower values \cite{shcherbakov:12a}.

Another particularly exciting development, again aimed squarely at Sgr~A$^*$ and the SMBH in the nearby elliptical galaxy M87, are the attempts to directly image the horizon-scale structures using mm- or submm-wave Very Long Baseline Interferometry (VLBI).  At these short radio wavelengths, the accretion flow is optically-thin to synchrotron self-absorption and hence we might hope to see the shadowing effect of the event horizon against the background radio emission \cite{falcke:00a}.   The properties of the shadow depend upon the spin (and spin-orientation) of the SMBH \cite{huang:07a,johannsen:10a}.  Great strides have already been made towards the goal of detecting these structures, with current interferometry being able to detect horizon scale structure but with an insufficient number of baselines to enable true image reconstruction \cite{doeleman:08a}.  In M87, current data have also resolved a compact ($10M$) structure possibly related to the jet-launching region \cite{doeleman:12a} which may, itself, probe SMBH spin as well as jet launching \cite{dexter:11a}. It will be exciting to watch these developments in the next few years.  

Throughout this review, we have assumed the validity of standard GR (and hence the Kerr metric) and have focused on attempted to extract the spin parameter.  A more ambitious problem is to use (electromagnetic) observations of accreting black holes to explore deviations from and/or extensions to GR \cite{bambi:09a,bambi:10a,johannsen:13a,bambi:13a,bambi:13b}.  A generic difficulty encountered in these studies is that, to lowest order, the deviations from GR are present in the quadrupole moment of the potential and hence are degenerate with the spin parameter.  Still, the implications are profound, and increasingly detailed studies of BH systems with electromagnetic and, eventually, gravitational wave probes remain our best hope for breaking GR.

The author thanks Laura Brenneman, Philip Cowperthwaite, Andrew Fabian, Anne Lohfink, Jon Miller, Rubens Reis, and Abdu Zoghbi for stimulating conservations that have helped formulate the discussion presented here.   The author also gratefully acknowledges the hospitality of the Institute of Astronomy at the National Central University (Jhongli, Taiwan) during which much of this review was written, as well as funding from NASA under grants NNX10AE41G (Astrophysics Theory), NNX10AR31G (Suzaku Guest Observer Program) and NNX12AE13G (Astrophysics Data Analysis).

\section*{References}

\bibliographystyle{unsrt}
\bibliography{blrgs,seyferts,chris}

\begin{thebibliography}{10}

\bibitem{kormendy:95a}
J.~{Kormendy} and D.~{Richstone}.
\newblock {Inward Bound---The Search For Supermassive Black Holes In Galactic
  Nuclei}.
\newblock {\em \araa}, 33:581, 1995.

\bibitem{rees:84a}
M.~J. {Rees}.
\newblock {Black Hole Models for Active Galactic Nuclei}.
\newblock {\em \araa}, 22:471--506, 1984.

\bibitem{benson:10a}
A.~J. {Benson}.
\newblock {Galaxy formation theory}.
\newblock {\em \physrep}, 495:33--86, October 2010.

\bibitem{fabian:12b}
A.~C. {Fabian}.
\newblock {Observational Evidence of Active Galactic Nuclei Feedback}.
\newblock {\em \araa}, 50:455--489, September 2012.

\bibitem{berti:13a}
E.~{Berti}.
\newblock {Astrophysical Black Holes as Natural Laboratories for Fundamental
  Physics and Strong-Field Gravity}.
\newblock {\em Brazilian Journal of Physics}, May 2013.

\bibitem{miller:07a}
J.~M. {Miller}.
\newblock {Relativistic X-Ray Lines from the Inner Accretion Disks Around Black
  Holes}.
\newblock {\em \araa}, 45:441--479, September 2007.

\bibitem{mcclintock:13a}
J.~E. {McClintock}, R.~{Narayan}, and J.~F. {Steiner}.
\newblock {Black Hole Spin via Continuum Fitting and the Role of Spin in
  Powering Transient Jets}.
\newblock {\em ArXiv e-prints}, March 2013.

\bibitem{moderski:96a}
R.~{Moderski} and M.~{Sikora}.
\newblock {On black hole evolution in active galactic nuclei}.
\newblock {\em \mnras}, 283:854--864, December 1996.

\bibitem{volonteri:05a}
M.~{Volonteri}, P.~{Madau}, E.~{Quataert}, and M.~J. {Rees}.
\newblock {The Distribution and Cosmic Evolution of Massive Black Hole Spins}.
\newblock {\em \apj}, 620:69--77, February 2005.

\bibitem{blandford:77a}
R.~D. {Blandford} and R.~L. {Znajek}.
\newblock {Electromagnetic extraction of energy from Kerr black holes}.
\newblock {\em \mnras}, 179:433--456, May 1977.

\bibitem{mckinney:12a}
J.~C. {McKinney}, A.~{Tchekhovskoy}, and R.~D. {Blandford}.
\newblock {General relativistic magnetohydrodynamic simulations of magnetically
  choked accretion flows around black holes}.
\newblock {\em \mnras}, 423:3083--3117, July 2012.

\bibitem{johannsen:10a}
T.~{Johannsen} and D.~{Psaltis}.
\newblock {Testing the No-hair Theorem with Observations in the Electromagnetic
  Spectrum. II. Black Hole Images}.
\newblock {\em \apj}, 718:446--454, July 2010.

\bibitem{brito:13a}
R.~{Brito}, V.~{Cardoso}, and P.~{Pani}.
\newblock {Massive spin-2 fields on black hole spacetimes: Instability of the
  Schwarzschild and Kerr solutions and bounds on graviton mass}.
\newblock {\em ArXiv e-prints}, April 2013.

\bibitem{brito:13b}
R.~{Brito}, V.~{Cardoso}, and P.~{Pani}.
\newblock {Partially massless gravitons do not destroy general relativity black
  holes}.
\newblock {\em ArXiv e-prints}, June 2013.

\bibitem{2013arXiv1305.5720C}
T.~e. {Consortium}, {:}, P.~A. {Seoane}, S.~{Aoudia}, H.~{Audley}, G.~{Auger},
  S.~{Babak}, J.~{Baker}, E.~{Barausse}, S.~{Barke}, M.~{Bassan},
  V.~{Beckmann}, M.~{Benacquista}, P.~L. {Bender}, E.~{Berti},
  P.~{Bin{\'e}truy}, J.~{Bogenstahl}, C.~{Bonvin}, D.~{Bortoluzzi}, N.~C.
  {Brause}, J.~{Brossard}, S.~{Buchman}, I.~{Bykov}, J.~{Camp}, C.~{Caprini},
  A.~{Cavalleri}, M.~{Cerdonio}, G.~{Ciani}, M.~{Colpi}, G.~{Congedo},
  J.~{Conklin}, N.~{Cornish}, K.~{Danzmann}, G.~{de Vine}, D.~{DeBra}, M.~{Dewi
  Freitag}, L.~{Di Fiore}, M.~{Diaz Aguilo}, I.~{Diepholz}, R.~{Dolesi},
  M.~{Dotti}, G.~{Fern{\'a}ndez Barranco}, L.~{Ferraioli}, V.~{Ferroni},
  N.~{Finetti}, E.~{Fitzsimons}, J.~{Gair}, F.~{Galeazzi}, A.~{Garcia},
  O.~{Gerberding}, L.~{Gesa}, D.~{Giardini}, F.~{Gibert}, C.~{Grimani},
  P.~{Groot}, F.~{Guzman Cervantes}, Z.~{Haiman}, H.~{Halloin}, G.~{Heinzel},
  M.~{Hewitson}, C.~{Hogan}, D.~{Holz}, A.~{Hornstrup}, D.~{Hoyland}, C.~D.
  {Hoyle}, M.~{Hueller}, S.~{Hughes}, P.~{Jetzer}, V.~{Kalogera},
  N.~{Karnesis}, M.~{Kilic}, C.~{Killow}, W.~{Klipstein}, E.~{Kochkina},
  N.~{Korsakova}, A.~{Krolak}, S.~{Larson}, M.~{Lieser}, T.~{Littenberg},
  J.~{Livas}, I.~{Lloro}, D.~{Mance}, P.~{Madau}, P.~{Maghami}, C.~{Mahrdt},
  T.~{Marsh}, I.~{Mateos}, L.~{Mayer}, D.~{McClelland}, K.~{McKenzie},
  S.~{McWilliams}, S.~{Merkowitz}, C.~{Miller}, S.~{Mitryk}, J.~{Moerschell},
  S.~{Mohanty}, A.~{Monsky}, G.~{Mueller}, V.~{M{\"u}ller}, G.~{Nelemans},
  D.~{Nicolodi}, S.~{Nissanke}, M.~{Nofrarias}, K.~{Numata}, F.~{Ohme},
  M.~{Otto}, M.~{Perreur-Lloyd}, A.~{Petiteau}, E.~S. {Phinney}, E.~{Plagnol},
  S.~{Pollack}, E.~{Porter}, P.~{Prat}, A.~{Preston}, T.~{Prince}, J.~{Reiche},
  D.~{Richstone}, D.~{Robertson}, E.~M. {Rossi}, S.~{Rosswog}, L.~{Rubbo},
  A.~{Ruiter}, J.~{Sanjuan}, B.~S. {Sathyaprakash}, S.~{Schlamminger},
  B.~{Schutz}, D.~{Sch{\"u}tze}, A.~{Sesana}, D.~{Shaddock}, S.~{Shah},
  B.~{Sheard}, C.~F. {Sopuerta}, A.~{Spector}, R.~{Spero}, R.~{Stanga},
  R.~{Stebbins}, G.~{Stede}, F.~{Steier}, T.~{Sumner}, K.-X. {Sun},
  A.~{Sutton}, T.~{Tanaka}, D.~{Tanner}, I.~{Thorpe}, M.~{Tr{\"o}bs},
  M.~{Tinto}, H.-B. {Tu}, M.~{Vallisneri}, D.~{Vetrugno}, S.~{Vitale},
  M.~{Volonteri}, V.~{Wand}, Y.~{Wang}, G.~{Wanner}, H.~{Ward}, B.~{Ware},
  P.~{Wass}, W.~J. {Weber}, Y.~{Yu}, N.~{Yunes}, and P.~{Zweifel}.
\newblock {The Gravitational Universe}.
\newblock {\em ArXiv e-prints}, May 2013.

\bibitem{kerr:63a}
R.~P. {Kerr}.
\newblock {Gravitational Field of a Spinning Mass as an Example of
  Algebraically Special Metrics}.
\newblock {\em Physical Review Letters}, 11:237--238, September 1963.

\bibitem{boyer:67a}
R.~H. {Boyer} and R.~W. {Lindquist}.
\newblock {Maximal Analytic Extension of the Kerr Metric}.
\newblock {\em Journal of Mathematical Physics}, 8:265--281, February 1967.

\bibitem{penrose:69a}
R.~Penrose.
\newblock Gravitational collapse: The role of general relativity.
\newblock {\em Riv. Nuovo Cimento}, 1:252--276, 1969.

\bibitem{mtw:73}
C.W. Misner, K.S. Thorne, and J.A. Wheeler.
\newblock {\em Gravitation}.
\newblock W.H. Freeman, San Fransisco, 1973.

\bibitem{gou:11a}
L.~{Gou}, J.~E. {McClintock}, M.~J. {Reid}, J.~A. {Orosz}, J.~F. {Steiner},
  R.~{Narayan}, J.~{Xiang}, R.~A. {Remillard}, K.~A. {Arnaud}, and S.~W.
  {Davis}.
\newblock {The Extreme Spin of the Black Hole in Cygnus X-1}.
\newblock {\em \apj}, 742:85, December 2011.

\bibitem{shakura:73a}
N.~I. {Shakura} and R.~A. {Sunyaev}.
\newblock {Black holes in binary systems. Observational appearance.}
\newblock {\em \aap}, 24:337--355, 1973.

\bibitem{balbus:91a}
S.~A. {Balbus} and J.~F. {Hawley}.
\newblock {A powerful local shear instability in weakly magnetized disks. I -
  Linear analysis. II - Nonlinear evolution}.
\newblock {\em \apj}, 376:214--233, July 1991.

\bibitem{reynolds:08a}
C.~S. {Reynolds} and A.~C. {Fabian}.
\newblock {Broad Iron-K{$\alpha$} Emission Lines as a Diagnostic of Black Hole
  Spin}.
\newblock {\em \apj}, 675:1048--1056, March 2008.

\bibitem{penna:10a}
R.~F. {Penna}, J.~C. {McKinney}, R.~{Narayan}, A.~{Tchekhovskoy}, R.~{Shafee},
  and J.~E. {McClintock}.
\newblock {Simulations of magnetized discs around black holes: effects of black
  hole spin, disc thickness and magnetic field geometry}.
\newblock {\em \mnras}, 408:752--782, October 2010.

\bibitem{bardeen:75a}
J.~M. {Bardeen} and J.~A. {Petterson}.
\newblock {The Lense-Thirring Effect and Accretion Disks around Kerr Black
  Holes}.
\newblock {\em \apjl}, 195:L65, January 1975.

\bibitem{lightman:88a}
A.~P. {Lightman} and T.~R. {White}.
\newblock {Effects of cold matter in active galactic nuclei - A broad hump in
  the X-ray spectra}.
\newblock {\em \apj}, 335:57--66, December 1988.

\bibitem{george:91a}
I.~M. {George} and A.~C. {Fabian}.
\newblock {X-ray reflection from cold matter in active galactic nuclei and
  X-ray binaries}.
\newblock {\em \mnras}, 249:352--367, March 1991.

\bibitem{ross:05a}
R.~R. {Ross} and A.~C. {Fabian}.
\newblock {A comprehensive range of X-ray ionized-reflection models}.
\newblock {\em \mnras}, 358:211--216, March 2005.

\bibitem{garcia:13a}
J.~{Garc{\'{\i}}a}, T.~{Dauser}, C.~S. {Reynolds}, T.~R. {Kallman}, J.~E.
  {McClintock}, J.~{Wilms}, and W.~{Eikmann}.
\newblock {X-Ray Reflected Spectra from Accretion Disk Models. III. A Complete
  Grid of Ionized Reflection Calculations}.
\newblock {\em \apj}, 768:146, May 2013.

\bibitem{fabian:89a}
A.~C. {Fabian}, M.~J. {Rees}, L.~{Stella}, and N.~E. {White}.
\newblock {X-ray fluorescence from the inner disc in Cygnus X-1}.
\newblock {\em \mnras}, 238:729--736, May 1989.

\bibitem{dauser:10a}
T.~{Dauser}, J.~{Wilms}, C.~S. {Reynolds}, and L.~W. {Brenneman}.
\newblock {Broad emission lines for a negatively spinning black hole}.
\newblock {\em \mnras}, 409:1534--1540, December 2010.

\bibitem{cunningham:75a}
C.~T. {Cunningham}.
\newblock {The effects of redshifts and focusing on the spectrum of an
  accretion disk around a Kerr black hole}.
\newblock {\em \apj}, 202:788--802, December 1975.

\bibitem{laor:91a}
A.~{Laor}.
\newblock {Line profiles from a disk around a rotating black hole}.
\newblock {\em \apj}, 376:90--94, July 1991.

\bibitem{brenneman:06a}
L.~W. {Brenneman} and C.~S. {Reynolds}.
\newblock {Constraining Black Hole Spin via X-Ray Spectroscopy}.
\newblock {\em \apj}, 652:1028--1043, December 2006.

\bibitem{fabian:02a}
A.~C. {Fabian}, S.~{Vaughan}, K.~{Nandra}, K.~{Iwasawa}, D.~R. {Ballantyne},
  J.~C. {Lee}, A.~{De Rosa}, A.~{Turner}, and A.~J. {Young}.
\newblock {A long hard look at MCG-6-30-15 with XMM-Newton}.
\newblock {\em \mnras}, 335:L1--L5, September 2002.

\bibitem{brenneman:11a}
L.~W. {Brenneman}, C.~S. {Reynolds}, M.~A. {Nowak}, R.~C. {Reis}, M.~{Trippe},
  A.~C. {Fabian}, K.~{Iwasawa}, J.~C. {Lee}, J.~M. {Miller}, R.~F. {Mushotzky},
  K.~{Nandra}, and M.~{Volonteri}.
\newblock {The Spin of the Supermassive Black Hole in NGC 3783}.
\newblock {\em \apj}, 736:103, August 2011.

\bibitem{fabian:12c}
A.~C. {Fabian}, D.~R. {Wilkins}, J.~M. {Miller}, R.~C. {Reis}, C.~S.
  {Reynolds}, E.~M. {Cackett}, M.~A. {Nowak}, G.~G. {Pooley}, K.~{Pottschmidt},
  J.~S. {Sanders}, R.~R. {Ross}, and J.~{Wilms}.
\newblock {On the determination of the spin of the black hole in Cyg X-1 from
  X-ray reflection spectra}.
\newblock {\em \mnras}, 424:217--223, July 2012.

\bibitem{wilkins:11a}
D.~R. {Wilkins} and A.~C. {Fabian}.
\newblock {Determination of the X-ray reflection emissivity profile of 1H
  0707-495}.
\newblock {\em \mnras}, 414:1269--1277, June 2011.

\bibitem{reis:12a}
R.~C. {Reis}, A.~C. {Fabian}, C.~S. {Reynolds}, L.~W. {Brenneman}, D.~J.
  {Walton}, M.~{Trippe}, J.~M. {Miller}, R.~F. {Mushotzky}, and M.~A. {Nowak}.
\newblock {X-Ray Spectral Variability in NGC 3783}.
\newblock {\em \apj}, 745:93, January 2012.

\bibitem{halpern:84a}
J.~P. {Halpern}.
\newblock {Variable X-ray absorption in the QSO MR 2251 - 178}.
\newblock {\em \apj}, 281:90--94, June 1984.

\bibitem{reynolds:97a}
C.~S. {Reynolds}.
\newblock {An X-ray spectral study of 24 type 1 active galactic nuclei}.
\newblock {\em \mnras}, 286:513--537, April 1997.

\bibitem{netzer:03a}
H.~{Netzer}, S.~{Kaspi}, E.~{Behar}, W.~N. {Brandt}, D.~{Chelouche}, I.~M.
  {George}, D.~M. {Crenshaw}, J.~R. {Gabel}, F.~W. {Hamann}, S.~B. {Kraemer},
  G.~A. {Kriss}, K.~{Nandra}, B.~M. {Peterson}, J.~C. {Shields}, and T.~J.
  {Turner}.
\newblock {The Ionized Gas and Nuclear Environment in NGC 3783. IV. Variability
  and Modeling of the 900 Kilosecond Chandra Spectrum}.
\newblock {\em \apj}, 599:933--948, December 2003.

\bibitem{mckernan:07a}
B.~{McKernan}, T.~{Yaqoob}, and C.~S. {Reynolds}.
\newblock {A soft X-ray study of type I active galactic nuclei observed with
  Chandra high-energy transmission grating spectrometer}.
\newblock {\em \mnras}, 379:1359--1372, August 2007.

\bibitem{kallman:01a}
T.~{Kallman} and M.~{Bautista}.
\newblock {Photoionization and High-Density Gas}.
\newblock {\em \apjs}, 133:221--253, March 2001.

\bibitem{reynolds:12a}
C.~S. {Reynolds}, L.~W. {Brenneman}, A.~M. {Lohfink}, M.~L. {Trippe}, J.~M.
  {Miller}, A.~C. {Fabian}, and M.~A. {Nowak}.
\newblock {A Monte Carlo Markov Chain Based Investigation of Black Hole Spin in
  the Active Galaxy NGC 3783}.
\newblock {\em \apj}, 755:88, August 2012.

\bibitem{miller:08a}
L.~{Miller}, T.~J. {Turner}, and J.~N. {Reeves}.
\newblock {An absorption origin for the X-ray spectral variability of
  MCG-6-30-15}.
\newblock {\em \aap}, 483:437--452, May 2008.

\bibitem{morgan:12a}
C.~W. {Morgan}, L.~J. {Hainline}, B.~{Chen}, M.~{Tewes}, C.~S. {Kochanek},
  X.~{Dai}, S.~{Kozlowski}, J.~A. {Blackburne}, A.~M. {Mosquera}, G.~{Chartas},
  F.~{Courbin}, and G.~{Meylan}.
\newblock {Further Evidence that Quasar X-Ray Emitting Regions are Compact:
  X-Ray and Optical Microlensing in the Lensed Quasar Q J0158-4325}.
\newblock {\em \apj}, 756:52, September 2012.

\bibitem{risaliti:13a}
G.~{Risaliti}, F.~A. {Harrison}, K.~K. {Madsen}, D.~J. {Walton}, S.~E. {Boggs},
  F.~E. {Christensen}, W.~W. {Craig}, B.~W. {Grefenstette}, C.~J. {Hailey},
  E.~{Nardini}, D.~{Stern}, and W.~W. {Zhang}.
\newblock {A rapidly spinning supermassive black hole at the centre of
  NGC1365}.
\newblock {\em \nat}, 494:449--451, February 2013.

\bibitem{marscher:02a}
A.~P. {Marscher}, S.~G. {Jorstad}, J.-L. {G{\'o}mez}, M.~F. {Aller},
  H.~{Ter{\"a}sranta}, M.~L. {Lister}, and A.~M. {Stirling}.
\newblock {Observational evidence for the accretion-disk origin for a radio jet
  in an active galaxy}.
\newblock {\em \nat}, 417:625--627, June 2002.

\bibitem{chatterjee:09a}
R.~{Chatterjee}, A.~P. {Marscher}, S.~G. {Jorstad}, A.~R. {Olmstead}, I.~M.
  {McHardy}, M.~F. {Aller}, H.~D. {Aller}, A.~{L{\"a}hteenm{\"a}ki},
  M.~{Tornikoski}, T.~{Hovatta}, K.~{Marshall}, H.~R. {Miller}, W.~T. {Ryle},
  B.~{Chicka}, A.~J. {Benker}, M.~C. {Bottorff}, D.~{Brokofsky}, J.~S.
  {Campbell}, T.~S. {Chonis}, C.~M. {Gaskell}, E.~R. {Gaynullina}, K.~N.
  {Grankin}, C.~H. {Hedrick}, M.~A. {Ibrahimov}, E.~S. {Klimek}, A.~K. {Kruse},
  S.~{Masatoshi}, T.~R. {Miller}, H.-J. {Pan}, E.~A. {Petersen}, B.~W.
  {Peterson}, Z.~{Shen}, D.~V. {Strel'nikov}, J.~{Tao}, A.~E. {Watkins}, and
  K.~{Wheeler}.
\newblock {Disk-Jet Connection in the Radio Galaxy 3C 120}.
\newblock {\em \apj}, 704:1689--1703, October 2009.

\bibitem{chatterjee:11a}
R.~{Chatterjee}, A.~P. {Marscher}, S.~G. {Jorstad}, A.~{Markowitz},
  E.~{Rivers}, R.~E. {Rothschild}, I.~M. {McHardy}, M.~F. {Aller}, H.~D.
  {Aller}, A.~{L{\"a}hteenm{\"a}ki}, M.~{Tornikoski}, B.~{Harrison},
  I.~{Agudo}, J.~L. {G{\'o}mez}, B.~W. {Taylor}, and M.~{Gurwell}.
\newblock {Connection Between the Accretion Disk and Jet in the Radio Galaxy 3C
  111}.
\newblock {\em \apj}, 734:43, June 2011.

\bibitem{lohfink:13a}
A.~M. {Lohfink}, C.~S. {Reynolds}, S.~G. {Jorstad}, A.~P. {Marscher}, E.~D.
  {Miller}, H.~{Aller}, M.~F. {Aller}, L.~W. {Brenneman}, A.~C. {Fabian}, J.~M.
  {Miller}, R.~F. {Mushotzky}, M.~A. {Nowak}, and F.~{Tombesi}.
\newblock {An X-Ray View of the Jet-Cycle in the Radio Loud AGN 3C120}.
\newblock {\em ArXiv e-prints}, May 2013.

\bibitem{lohfink:12b}
A.~M. {Lohfink}, C.~S. {Reynolds}, J.~M. {Miller}, L.~W. {Brenneman}, R.~F.
  {Mushotzky}, M.~A. {Nowak}, and A.~C. {Fabian}.
\newblock {The Black Hole Spin and Soft X-Ray Excess of the Luminous Seyfert
  Galaxy Fairall 9}.
\newblock {\em \apj}, 758:67, October 2012.

\bibitem{walton:13a}
D.~J. {Walton}, E.~{Nardini}, A.~C. {Fabian}, L.~C. {Gallo}, and R.~C. {Reis}.
\newblock {Suzaku observations of `bare' active galactic nuclei}.
\newblock {\em \mnras}, 428:2901--2920, February 2013.

\bibitem{patrick:12a}
A.~R. {Patrick}, J.~N. {Reeves}, D.~{Porquet}, A.~G. {Markowitz}, V.~{Braito},
  and A.~P. {Lobban}.
\newblock {A Suzaku survey of Fe K lines in Seyfert 1 active galactic nuclei}.
\newblock {\em \mnras}, 426:2522--2565, November 2012.

\bibitem{reynolds:13a}
C.~S. {Reynolds}.
\newblock {Measuring Black Hole Spin using X-ray Reflection Spectroscopy}.
\newblock {\em ArXiv e-prints}, February 2013.

\bibitem{king:04a}
A.~R. {King}, J.~E. {Pringle}, R.~G. {West}, and M.~{Livio}.
\newblock {Variability in black hole accretion discs}.
\newblock {\em \mnras}, 348:111--122, February 2004.

\bibitem{cowperthwaite:12a}
P.~S. {Cowperthwaite} and C.~S. {Reynolds}.
\newblock {The Central Engine Structure of 3C120: Evidence for a Retrograde
  Black Hole or a Refilling Accretion Disk}.
\newblock {\em \apjl}, 752:L21, June 2012.

\bibitem{reynolds:99a}
C.~S. {Reynolds}, A.~J. {Young}, M.~C. {Begelman}, and A.~C. {Fabian}.
\newblock {X-Ray Iron Line Reverberation from Black Hole Accretion Disks}.
\newblock {\em \apj}, 514:164--179, March 1999.

\bibitem{shapiro:64a}
I.~I. {Shapiro}.
\newblock {Fourth Test of General Relativity}.
\newblock {\em Physical Review Letters}, 13:789--791, December 1964.

\bibitem{zoghbi:12a}
A.~{Zoghbi}, A.~C. {Fabian}, C.~S. {Reynolds}, and E.~M. {Cackett}.
\newblock {Relativistic iron K X-ray reverberation in NGC 4151}.
\newblock {\em \mnras}, 422:129--134, May 2012.

\bibitem{fabian:09a}
A.~C. {Fabian}, A.~{Zoghbi}, R.~R. {Ross}, P.~{Uttley}, L.~C. {Gallo}, W.~N.
  {Brandt}, A.~J. {Blustin}, T.~{Boller}, M.~D. {Caballero-Garcia},
  J.~{Larsson}, J.~M. {Miller}, G.~{Miniutti}, G.~{Ponti}, R.~C. {Reis}, C.~S.
  {Reynolds}, Y.~{Tanaka}, and A.~J. {Young}.
\newblock {Broad line emission from iron K- and L-shell transitions in the
  active galaxy 1H0707-495}.
\newblock {\em \nat}, 459:540--542, May 2009.

\bibitem{zoghbi:10a}
A.~{Zoghbi}, A.~C. {Fabian}, P.~{Uttley}, G.~{Miniutti}, L.~C. {Gallo}, C.~S.
  {Reynolds}, J.~M. {Miller}, and G.~{Ponti}.
\newblock {Broad iron L line and X-ray reverberation in 1H0707-495}.
\newblock {\em \mnras}, 401:2419--2432, February 2010.

\bibitem{demarco:13a}
B.~{De Marco}, G.~{Ponti}, M.~{Cappi}, M.~{Dadina}, P.~{Uttley}, E.~M.
  {Cackett}, A.~C. {Fabian}, and G.~{Miniutti}.
\newblock {Discovery of a relation between black hole mass and soft X-ray time
  lags in active galactic nuclei}.
\newblock {\em \mnras}, 431:2441--2452, May 2013.

\bibitem{zoghbi:13a}
A.~{Zoghbi}, C.~{Reynolds}, E.~M. {Cackett}, G.~{Miniutti}, E.~{Kara}, and
  A.~C. {Fabian}.
\newblock {Discovery of Fe K{$\alpha$} X-Ray Reverberation around the Black
  Holes in MCG-5-23-16 and NGC 7314}.
\newblock {\em \apj}, 767:121, April 2013.

\bibitem{kara:13a}
E.~{Kara}, A.~C. {Fabian}, E.~M. {Cackett}, P.~{Uttley}, D.~R. {Wilkins}, and
  A.~{Zoghbi}.
\newblock {Discovery of high-frequency iron K lags in Ark 564 and Mrk 335}.
\newblock {\em ArXiv e-prints}, June 2013.

\bibitem{chartas:09a}
G.~{Chartas}, C.~S. {Kochanek}, X.~{Dai}, S.~{Poindexter}, and G.~{Garmire}.
\newblock {X-Ray Microlensing in RXJ1131-1231 and HE1104-1805}.
\newblock {\em \apj}, 693:174--185, March 2009.

\bibitem{dai:10a}
X.~{Dai}, C.~S. {Kochanek}, G.~{Chartas}, S.~{Koz{\l}owski}, C.~W. {Morgan},
  G.~{Garmire}, and E.~{Agol}.
\newblock {The Sizes of the X-ray and Optical Emission Regions of RXJ
  1131-1231}.
\newblock {\em \apj}, 709:278--285, January 2010.

\bibitem{davis:11a}
S.~W. {Davis} and A.~{Laor}.
\newblock {The Radiative Efficiency of Accretion Flows in Individual Active
  Galactic Nuclei}.
\newblock {\em \apj}, 728:98, February 2011.

\bibitem{daly:11a}
R.~A. {Daly}.
\newblock {Estimates of black hole spin properties of 55 sources}.
\newblock {\em \mnras}, 414:1253--1262, June 2011.

\bibitem{rees:82a}
M.~J. {Rees}, M.~C. {Begelman}, R.~D. {Blandford}, and E.~S. {Phinney}.
\newblock {Ion-supported tori and the origin of radio jets}.
\newblock {\em \nat}, 295:17--21, January 1982.

\bibitem{narayan:95a}
R.~{Narayan}, I.~{Yi}, and R.~{Mahadevan}.
\newblock {Explaining the spectrum of Sagittarius A$^{*}$ with a model of an
  accreting black hole}.
\newblock {\em \nat}, 374:623--625, April 1995.

\bibitem{moscibrodzka:09a}
M.~{Mo{\'s}cibrodzka}, C.~F. {Gammie}, J.~C. {Dolence}, H.~{Shiokawa}, and
  P.~K. {Leung}.
\newblock {Radiative Models of SGR A* from GRMHD Simulations}.
\newblock {\em \apj}, 706:497--507, November 2009.

\bibitem{dexter:11a}
J.~{Dexter}, J.~C. {McKinney}, and E.~{Agol}.
\newblock {The size of the jet launching region in M87}.
\newblock {\em \mnras}, 421:1517--1528, April 2012.

\bibitem{shcherbakov:12a}
R.~V. {Shcherbakov}, R.~F. {Penna}, and J.~C. {McKinney}.
\newblock {Sagittarius A* Accretion Flow and Black Hole Parameters from General
  Relativistic Dynamical and Polarized Radiative Modeling}.
\newblock {\em \apj}, 755:133, August 2012.

\bibitem{falcke:00a}
H.~{Falcke}, F.~{Melia}, and E.~{Agol}.
\newblock {Viewing the Shadow of the Black Hole at the Galactic Center}.
\newblock {\em \apjl}, 528:L13--L16, January 2000.

\bibitem{huang:07a}
L.~{Huang}, M.~{Cai}, Z.-Q. {Shen}, and F.~{Yuan}.
\newblock {Black hole shadow image and visibility analysis of Sagittarius A*}.
\newblock {\em \mnras}, 379:833--840, August 2007.

\bibitem{doeleman:08a}
S.~S. {Doeleman}, J.~{Weintroub}, A.~E.~E. {Rogers}, R.~{Plambeck},
  R.~{Freund}, R.~P.~J. {Tilanus}, P.~{Friberg}, L.~M. {Ziurys}, J.~M. {Moran},
  B.~{Corey}, K.~H. {Young}, D.~L. {Smythe}, M.~{Titus}, D.~P. {Marrone}, R.~J.
  {Cappallo}, D.~C.-J. {Bock}, G.~C. {Bower}, R.~{Chamberlin}, G.~R. {Davis},
  T.~P. {Krichbaum}, J.~{Lamb}, H.~{Maness}, A.~E. {Niell}, A.~{Roy},
  P.~{Strittmatter}, D.~{Werthimer}, A.~R. {Whitney}, and D.~{Woody}.
\newblock {Event-horizon-scale structure in the supermassive black hole
  candidate at the Galactic Centre}.
\newblock {\em \nat}, 455:78--80, September 2008.

\bibitem{doeleman:12a}
S.~S. {Doeleman}, V.~L. {Fish}, D.~E. {Schenck}, C.~{Beaudoin}, R.~{Blundell},
  G.~C. {Bower}, A.~E. {Broderick}, R.~{Chamberlin}, R.~{Freund}, P.~{Friberg},
  M.~A. {Gurwell}, P.~T.~P. {Ho}, M.~{Honma}, M.~{Inoue}, T.~P. {Krichbaum},
  J.~{Lamb}, A.~{Loeb}, C.~{Lonsdale}, D.~P. {Marrone}, J.~M. {Moran},
  T.~{Oyama}, R.~{Plambeck}, R.~A. {Primiani}, A.~E.~E. {Rogers}, D.~L.
  {Smythe}, J.~{SooHoo}, P.~{Strittmatter}, R.~P.~J. {Tilanus}, M.~{Titus},
  J.~{Weintroub}, M.~{Wright}, K.~H. {Young}, and L.~M. {Ziurys}.
\newblock {Jet Launching Structure Resolved Near the Supermassive Black Hole in
  M87}.
\newblock {\em ArXiv e-prints}, October 2012.

\bibitem{bambi:09a}
C.~{Bambi} and K.~{Freese}.
\newblock {Apparent shape of super-spinning black holes}.
\newblock {\em \prd}, 79(4):043002, February 2009.

\bibitem{bambi:10a}
C.~{Bambi} and N.~{Yoshida}.
\newblock {Shape and position of the shadow in the {$\delta$} = 2
  Tomimatsu-Sato spacetime}.
\newblock {\em Classical and Quantum Gravity}, 27(20):205006, October 2010.

\bibitem{johannsen:13a}
T.~{Johannsen} and D.~{Psaltis}.
\newblock {Testing the No-hair Theorem with Observations in the Electromagnetic
  Spectrum. IV. Relativistically Broadened Iron Lines}.
\newblock {\em \apj}, 773:57, August 2013.

\bibitem{bambi:13a}
C.~{Bambi}.
\newblock {Testing the space-time geometry around black hole candidates with
  the analysis of the broad K{$\alpha$} iron line}.
\newblock {\em \prd}, 87(2):023007, January 2013.

\bibitem{bambi:13b}
C.~{Bambi} and D.~{Malafarina}.
\newblock {K\$$\backslash$alpha\$ iron line profile from accretion disks around
  regular and singular exotic compact objects}.
\newblock {\em ArXiv e-prints}, July 2013.

\end{thebibliography}

\end{document}